\documentclass[pra,reprint]{revtex4-1}
\usepackage{graphicx}
\usepackage{enumitem}
\usepackage{amsmath}
\newcommand{\ket}[1]{|{#1}\rangle}

\newcommand{\braket}[1]{\langle{#1}}
\usepackage{amsfonts}

\usepackage[usenames,dvipsnames]{xcolor}
\usepackage[colorlinks=true,citecolor=Cerulean,linkcolor=RubineRed,urlcolor=Cerulean]{hyperref}

\begin{document}

\title{Exact edge, bulk and bound states of finite topological systems}

\author{Callum W. Duncan}
\email{cd130@hw.ac.uk}
\author{Patrik \"{O}hberg}
\author{Manuel Valiente}
\email{M.Valiente$\_$Cifuentes@hw.ac.uk}
\affiliation{SUPA, Institute of Photonics and Quantum Sciences,
Heriot-Watt University, Edinburgh EH14 4AS, United Kingdom}

\begin{abstract}
Finite topologically non-trivial systems are characterised, among many other unique properties, by the presence of bound states at their physical edges. These topological edge modes can be distinguished from usual Shockley waves energetically, as their energies remain finite and in-gap even when the boundaries of the system represent an effectively infinite and sharp energetic barrier. Theoretically, the existence of topological edge modes can be shown by means of the bulk-edge correspondence and topological invariants. On a clean one-dimensional lattice and reducible two-dimensional models, in either the commensurate or semi-infinite case, the edge modes can be essentially obtained analytically, as shown in [Phys. Rev. Lett. {\bf 71}, 3697 (1993)] and [Phys. Rev. A {\bf 89}, 023619 (2014)]. In this work, we put forward a method for obtaining the spectrum and wave functions of topological edge modes for arbitrary finite lattices, including the incommensurate case. A small number of parameters are easily determined numerically, with the form of the eigenstates remaining fully analytical. We also obtain the bulk modes in the finite system analytically and their associated eigenenergies, which lie within the infinite-size limit continuum. Our method is general and can be easily applied to obtain the properties of non-topological models and/or extended to include impurities. As an example, we consider a relevant case of an impurity located next to one edge of a one-dimensional system, equivalent to a softened boundary in a separable two-dimensional model. We show that a localised impurity can have a drastic effect on the original topological edge modes of the system. Using the periodic Harper and Hofstadter models to illustrate our method, we find that, on increasing the impurity strength, edge states can enter or exit the continuum, and a trivial Shockley state bound to the impurity may appear. The fate of the topological edge modes in the presence of impurities can be addressed by quenching the impurity strength. We find that at certain critical impurity strengths, the transition probability for a particle initially prepared in an edge mode to decay into the bulk exhibits discontinuities that mark the entry and exit points of edge modes from and into the bulk spectrum.                     
\end{abstract}

\maketitle

\section{Introduction}
The discovery of the quantum Hall \cite{Klitzing1980} and fractional quantum Hall effects \cite{Laughlin1981,Tsui1982}, and their subsequent characterisation in terms of topological invariants \cite{Thouless1982} lead to the establishment of a new paradigm that does not follow the premises of Landau's theory of symmetry breaking \cite{AndersonBook}. More recently, there has been a great surge of interest in topologically non-trivial systems due to the theoretical prediction and experimental realisation of time-reversal-symmetric topological insulators \cite{Bernevig2006,Fu2007,Hsieh2008,Konig2007,Zhang2009,Chen2009,Xia2009}. Topological systems can also be engineered out of equilibrium. In particular, time-periodic modulation of lattices can lead to quasi-static Floquet states with non-zero Chern numbers in the fastly-driven regime \cite{Goldman2014}, and chiral edge modes when boundaries are present. It is also possible to establish topological invariants -- winding numbers -- that go beyond the quasi-static case in slowly-driven Floquet systems \cite{Rudner2013}, and also exhibit topological edge modes, which have been experimentally observed with photonic lattices \cite{Mukherjee2017,Maczewsky2017}. The great generality of topological systems is further attested by experimental realisations in a variety of systems, including condensed matter \cite{Hsieh2008,Xia2009}, ultracold atoms \cite{Atala2013,Jotzu2014} and even mechanical systems \cite{Susstrunk2015}.  

One of the main features of symmetry-protected topological insulators is their robustness against weak local perturbations. For a symmetry protected topological state a closing of the bulk band gap is expected at any topological phase transition. This is according to the bulk-edge correspondence, as the gap must close for there to be a change in the topological invariant \cite{BernevigBook}. However, recently, topological phase transitions without closing the bulk band gap have become of interest \cite{Ezawa2013,Rachel2016}. These works consider Hamiltonians which are the sum of a topologically non-trivial Hamiltonian and a trivial {\it global} perturbation, whose bulks may even be isospectral, as in the case of the Su-Schrieffer-Heeger (SSH) and ionic Hubbard models \cite{Valiente2010}. Once the trivial global perturbation outweighs the topological unperturbed Hamiltonian, which by no means implies gap closing in general, it is obvious that the system will become non-topological. Therefore, these examples pose no threat to the robustness of topological phases against local perturbations, since these are global in nature and, quite trivially, do not require gap-closing for a topological transition to occur. 
There have also been many works studying the properties of topological insulators and their robustness against local perturbations and disorder \cite{Liu2009,Biswas2010,Lu2011,Black-Schaffer2012,Buchhold2012,Song2015,Groh2016,Zhou2016,Liberto2016,
Salerno2017,Malki2017,Balabanov2017,Lupke2017,Shiranzaei2017,Jorg2017}.

Besides the calculation of topological invariants, which give information about the number of topological modes \cite{Thouless1982,Goldman2013}, one may directly calculate the edge modes and their eigenenergies. While brute force numerical diagonalisation is always an option, there are some instances where edge modes have been obtained essentially analytically. For example, the Harper and Hofstadter models, investigated by Hatsugai \cite{Hatsugai1993} using the transfer matrix, and later on by H{\"u}gel and Paredes \cite{Hugel2014} using the extension of Bloch's theorem to complex quasi-momenta first developed in Ref.~\cite{Manuel2010}. These methods, unfortunately, only work in two particular situations, namely (i) on a semi-infinite lattice and (ii) on a commensurate lattice, i.e. with a number of lattice sites $N$ that is an integer multiple of the spatial period $\tau$ of the Hamiltonian, minus one ($N=n\tau -1$, with $n\in \mathbb{Z}_+$). The possibility of dealing with the last case is an immediate consequence of Bloch's theorem and case (i). Exact helical edge states have also been constructed from an ansatz in the quantum Hall and Bernevig-Hughes-Zhang models \cite{Konig2008,Mao2010,Mao2011}, under similar conditions. In addition, exact topological eigenstates have been found for stacked lattices \cite{Kunst2017,Kunst2018}, under certain conditions on the lattice structure, boundary conditions, commensurability and couplings between stacked layers.

The scattering matrix, or S-matrix, approach is a useful method to find bound states and their properties in open systems \cite{Taylor2006,Weinberg2015}. This approach has been of importance in the understanding of superconducting junctions \cite{Beenakker1991,Beenakker1992,Beenakker1994,Badiane2011,Pikulin2012}, and has also been utilised in the context of photonic crystals \cite{Xiao2014,Poshakinskiy2015,Shi2016}, the calculation of topological invariants \cite{Fulga2012} and the topic of quantum chaos in billiards \cite{Doron1992,Dietz1993}. Recently, the S-matrix method was used to solve for the Zak phase in one-dimensional tight-binding models \cite{Arkinstall2017}, bringing it closer to the models we will consider in this work.

For finite lattices with arbitrary parameters of general, incommensurate size, and for the aperiodic case -- which we shall call inhomogeneous from here on, while the clean case shall be called homogeneous -- where impurities are present, fully numerical solutions have been required. Moreover, the analytical derivation of bulk modes in the presence of any boundaries has not been addressed so far, not even in the semi-infinite and commensurate cases. There are a number of reasons why the knowledge of the exact form of all the eigenstates of topologically non-trivial systems is important, besides their theoretical relevance. First of all, purely numerical calculations cannot attach any quantum numbers to the eigenstates, except for their energy, a matter fully resolved by analytical methods when these exist. Moreover, the knowledge of the exact wave functions and their properties allows for much-needed simplifications when calculating interaction matrix elements in many-particle systems, or single-particle response functions that would otherwise have to be performed by means of either numerical brute force or Monte Carlo simulation.

In this work, we consider finite one-dimensional and separable two-dimensional topological lattice models. The Hamiltonians considered will be nearest-neighbour, tight-binding lattices with time-reversal symmetry preserved. We will not invoke the bulk-edge correspondence or topological invariants  to confirm the existence of topological edge states, for which there has been several previous theoretical works, see Refs.~\cite{Hasan2010,Delplace2011,Mong2011,
Rhim2017a,Rhim2017b} and references therein. Instead, we will set out a method for writing the full edge and bulk states analytically, for arbitrary (i.e. commensurate and incommensurate) system sizes. We will then extend our method to inhomogeneous systems which, for the sake of concreteness, will correspond to a single impurity located next to one edge of the system. Throughout the manuscript, we will use our method for particular models, namely the SSH and diagonal Harper models in one dimension, and the two-dimensional Hofstadter model on a square lattice. We find that a single impurity can have drastic effects on the edge modes of these systems when the impurity is of significant strength, i.e. on the order of the finite band gap. In the models we consider, we find that there can be several ``critical'' impurity interaction strengths at which edge modes may enter or exit the continua of bulk bands, while usual Shockley states \cite{Shockley1939} appear as the impurity strength is cranked up. Finally, we study dynamical transitions of edge modes into the bulk as the strength of the impurity is ramped. To do this, we consider an initially homogeneous system, with an edge mode as an initial state. We introduce a sudden increase to the impurity strength and find that the edge-to-bulk transition probabilities exhibit distinctive discontinuities at the different critical interaction strengths. The transition probabilities are not negligible for edge states bound to the boundary next to which the impurity is located, even at low and intermediate impurity strengths. The size of the discontinuity at the critical point where the initial edge state adiabatically connects to the bulk is quite large, and should be observable experimentally with ultracold atoms \cite{Meier2016,Schreiber2015,Miyake2013,Kennedy2015} or photonic lattices \cite{Lahini2009,Kraus2012}.    

\section{Hamiltonian of the system}
We shall consider here one-dimensional tight-binding lattice models whose Hamiltonians have non-trivial periodicities of $\tau$ lattice sites, $2\le \tau \in \mathbb{Z}_+$, and two-dimensional models that can be reduced to one dimension, resulting in single-parameter fibre Hamiltonians. In the latter case, we shall consider square lattices for simplicity, which result in one sublattice only, but note that other geometries, including graphene's non-interacting tight-binding Hamiltonian, can be reduced to one dimension, albeit with several sublattices \cite{Hatsugai2006}. Solving the 1D sublattice of graphene and obtaining the Zak phase has been considered in Ref.~\cite{Delplace2011}.

The general one-dimensional Hamiltonian we consider is of the form (we set the lattice spacing $d=1$)
\begin{align}
H = - \sum_x & \left( t_{x+1,x}\left(\xi,\tau_1\right) \hat{c}^{\dagger}_x \hat{c}_{x+1} + \mathrm{H.c.} \right) + \nonumber \\ & + \sum_x V_x\left(\lambda,\tau_2\right) \hat{c}^{\dagger}_x \hat{c}_{x}.\label{homogeneousHamiltonian}
\end{align}
Above, $\hat{c}_{x}$ ($ \hat{c}_{x}^{\dagger}$) annihilates (creates) a particle at site $x$ and we will take $x=x_0+1,\dots,L_s-1$, with open boundary conditions. We will consider the tunnelling coefficients $t_{x,x+1}$ and on-site potentials $V_x$ to have, in general, different periodicities $\tau_1$ and $\tau_2$, with the largest of them ($\tau$) being proportional to the smallest, and strengths controlled by dimensionless parameters $\xi$ and $\lambda$, respectively. We note that the eigenenergy $\epsilon_{\sigma}(k)$ associated with a Bloch state in band $\sigma$ at quasi-momentum $k$ ($\in (-\pi/\tau,\pi/\tau]$) satisfies $\epsilon_{\sigma}(k)=\epsilon_{\sigma}(-k)$, due to Hermiticity.  

\section{Eigenstates in the homogeneous case}
Here, we present a method to extract bulk and edge modes of topological models, with Hamiltonians of the form \eqref{homogeneousHamiltonian}, in a finite lattice under open boundary conditions, without any impurities (``homogeneous'' case). Although this is a particular case of the problem including impurities, it makes our discussion of the more general problem much clearer, as this is just a generalisation of the method. For later reference, we write a general single-particle eigenstate $\ket{\psi}$ of Hamiltonian (\ref{homogeneousHamiltonian}) in the form
\begin{equation}
  \ket{\psi}=\sum_{x=x_0+1}^{L_s-1}\psi(x)c_x^{\dagger}\ket{0},\label{generalEigenstate}
\end{equation}
where $\ket{0}$ is the vacuum of particles. Below, we first discuss the bulk modes of the system and then move on to discuss edge states.

\subsection{Bulk states}
By definition, we shall call bulk states in a finite lattice those eigenstates with associated eigenenergies lying in the infinite-size continuum.

In this limit ($-x_0,L_s\to +\infty$), all eigenstates have the form of Bloch states, i.e.
\begin{equation}
  \psi_{\sigma,k}(x)=e^{ikx}\phi_{\sigma,k}(x),
\end{equation}
where $\phi_{\sigma,k}(x+\tau)=\phi_{\sigma,k}(x)$ with $\sigma=1,2,\ldots,\tau$ the band index and $k\in (-\pi/\tau,\pi/\tau]$ the quasi-momentum.

Since the spectrum satisfies $\epsilon_{\sigma}(k)=\epsilon_{\sigma}(-k)$, we can freely superpose $\ket{\psi_{\sigma,k}}$ and $\ket{\psi_{\sigma,-k}}$ to form eigenstates. Since these are the only eigenstates with energies in the infinite-size continuum, the bulk states of 
the finite-size Hamiltonian (\ref{homogeneousHamiltonian}) must have the form  
\begin{align}
\psi_{\sigma,k} \left(x\right) = A \phi_{\sigma,+k} \left(x\right) e^{i k x} + B \phi_{\sigma,-k} \left(x\right) e^{-i k x}. \label{eq:BulkState}
\end{align}
Using the boundary conditions ($\psi_{\sigma,k}\left(x_0\right) = \psi_{\sigma,k}\left(L_s\right) = 0$) we obtain the general quantisation condition for $k$ 
\begin{align}
e^{2 i k \left(x_0 - L\right)} = \frac{\phi_{\sigma,+k} \left( L_s \right) \phi_{\sigma,-k} \left( x_0 \right)}{\phi_{\sigma,+k} \left( x_0 \right) \phi_{\sigma,-k} \left( L_s \right)}, \label{eq:BulkQuantisation}
\end{align}
and the coefficients in Eq.~(\ref{eq:BulkState}) are related as
\begin{equation}
  \frac{A}{B}=-\frac{\phi_{\sigma,-k}(x_0)}{\phi_{\sigma,+k}(x_0)}e^{-2ikx_0}=-\frac{\phi_{\sigma,-k}(L_s)}{\phi_{\sigma,+k}(L_s)}e^{-2ikL_s}.\label{coeff}
\end{equation}
We can now write the un-normalised bulk states in the form
\begin{align}
\psi_k \left(x\right) = \phi_{\sigma,+k} \left(x\right) e^{i k x} - \frac{\phi_{\sigma,+k} \left(L_s\right)}{\phi_{\sigma,-k} \left(L_s\right)} \phi_{\sigma,-k} \left(x\right) e^{-i k (x - 2 L_s)}, \label{eq:LeftBulk}
\end{align}
with the set of allowed quasi-momenta $k$ satisfying Eq.~(\ref{eq:BulkQuantisation}).


\subsection{Edge states}
When these exist, edge states are formally not different from usual bound states. Therefore, in the semi-infinite limit (either $x_0\to -\infty$ or $L_s\to \infty$), the edge states must have the form of 
\begin{align}
\psi\left(x\right) = \phi \left(x \right) \alpha^x,\label{eq:SingleAnsatz}
\end{align}
where $0 < |\alpha| < 1$, $x\ge x_0+1$ if $L_s\to \infty$ ($|\alpha|>1$, $x\le L_s-1$ if $-x_0\to \infty$). The Bloch functions in Eq.~(\ref{eq:SingleAnsatz}), $\phi \left(x \right)$, are found for a given $\alpha$ as if the edge states were bulk modes, i.e. by solving the Schr\"{o}dinger equation for $\psi$ in Eq.~(\ref{eq:SingleAnsatz}) with periodic boundary conditions for $\phi \left(x \right) = \phi \left(x + \tau \right)$. Note that $\alpha$ is allowed to be real or complex. In fact, its structure can be derived exactly from symmetry arguments. To see this, we write $\alpha=e^{ik}=e^{-\lambda+i\kappa}$, with $\lambda$ and $\kappa$ real. The corresponding energy is given by $E=\epsilon(i\lambda+\kappa)$. From the Schr{\"o}dinger equation for the Bloch states we see that $\epsilon(i\lambda+\kappa)=[\epsilon(i\lambda-\kappa))]^*$. The energies for the allowed values of $\alpha$ must be real, which implies that the allowed values for $\kappa$ are $\kappa=0,\pi/\tau$, i.e. vanishing quasi-momentum or its value at the edge of a band. We can see this using the periodicity property $\epsilon(k+2\pi/\tau)=\epsilon(k)$.

For the case of a system with $|L_s|,|x_0|<\infty$ the state with $|\alpha| > 1$ is normalizable. Hence, as long as the energy has $\alpha \rightarrow 1/\alpha$ symmetry -- guaranteed if $\epsilon(k)=\epsilon(-k)$ -- we can write the most general bound state wave function for finite $x_0$ and $L_s$ as
\begin{align}
\psi\left(x\right) = A \phi_+ \left(x \right) \alpha^x + B \phi_- \left(x \right) \alpha^{-x}.\label{eq:BoundState}
\end{align}
Note, that while the energy has the symmetry $\alpha \rightarrow 1/\alpha$, the Bloch states do not possess this symmetry in general. Hence, the notation of $\phi_{+\left(-\right)} \left(x \right)$ corresponding to the Bloch functions associated with the $\alpha \left(1/\alpha\right)$ bound states.

Imposing the open boundary conditions on the general bound state~(\ref{eq:BoundState}) implies the following transcendental equation for $\alpha$
\begin{align}
\alpha^{2(L_s-x_0)} = \frac{\phi_+ \left( x_0 \right) \phi_- \left( L_s \right)}{\phi_+ \left( L_s \right) \phi_- \left( x_0 \right)}.\label{eq:U0Solve}
\end{align}

The edge states can be written by applying one of the boundary conditions and then absorbing a factor into the normalisation. By applying the condition $\psi\left(L_s\right) = 0$ to Eq.~(\ref{eq:BoundState}) or applying $\psi\left(x_0=0\right)=0$, we obtain the un-normalised edge state to be of either of the two following forms
\begin{align}
  \psi \left( x \right) &= \phi_+ \left(x\right) \alpha^x - \frac{\phi_+\left(L_s\right)}{\phi_-\left(L_s\right)} \phi_- \left(x\right) \alpha^{2L_s-x}.\label{eq:LeftEdge}\\
  \psi \left( x \right) &= \phi_+ \left(x\right) \alpha^x - \frac{\phi_+\left(x_0\right)}{\phi_-\left(x_0\right)} \phi_- \left(x\right) \alpha^{2x_0-x}.\label{eq:RightEdge}
\end{align}
For concreteness, throughout this work we will make use of edge states written in the form of Eq.~(\ref{eq:LeftEdge}).

At this point, it is worth noting that though we write the solutions for the edge and bulk states separately, they are, of course, part of the same general solution. With the edge states being complex momentum solutions of the bulk state ansatz of Eq.~(\ref{eq:BulkState}), as used in the symmetry argument earlier in this section. In the simplest of cases, we can solve for all states using either the bulk or edge ansatz. For more complex systems it is conceptually simpler to separate into the bulk and edge ansatz and constrain the solutions appropriately to solve for the states.

\section{Eigenstates with an impurity}
Now we place an impurity of strength $U$, which can be arbitrarily attractive ($U<0$) or repulsive ($U>0$), on the lattice. We choose the location of the impurity to be $x=x_0+1$, i.e. the first lattice site of the finite system. This makes the solution slightly simpler than in the general case of arbitrary location since one only needs to deal with wave functions to the right of the impurity, but the solution can be obtained analogously in general. Moreover, this is the most physically relevant case, since it models edge softening. 

We begin by studying the edge states. These can still be written in the exact same form as in the homogeneous ($U=0$) case. However, Eq.~(\ref{eq:U0Solve}) for $\alpha$ is no longer valid, as the Bloch states do not take into account the presence of the impurity on the edge, they are solely dependent on the periodic nature of the lattice. To account for the impurity we are required to solve the Schr\"{o}dinger equation at $x=x_0+1$ in order to extract $\alpha$, as is usual for bound state problems on a lattice \cite{Valiente2008,Valiente2009,Valiente2010PRA}. The total Hamiltonian of the system now reads, in second quantisation,
\begin{equation}
  H=H_0+U\hat{c}_{x_0+1}^{\dagger}\hat{c}_{x_0+1},\label{FullHamiltonian}
\end{equation}
where $H_0$ is given by the impurity-free Hamiltonian of Eq.~(\ref{homogeneousHamiltonian}). We insert the general form of the edge modes with the open boundary conditions already implemented, Eq.~(\ref{eq:LeftEdge}), into the Schr{\"o}dinger equation for Hamiltonian (\ref{FullHamiltonian}) at $x=x_0+1$. We use the fact that $\psi(x_0)=0$ and obtain
\begin{equation}
  E=-t_{x_0+2,x_0+1}(\xi,\tau_1)\frac{\psi(x_0+2)}{\psi(x_0+1)}+V_{x_0+1}\left(\lambda,\tau_2\right)+U,
\end{equation}
where $\psi(x)$ is given by Eq.~(\ref{eq:LeftEdge}) for the topological edge state. The other equation for the eigenenergies $E$ is simply given by $E=\epsilon(i\lambda+\kappa)$, where $\alpha=e^{-\lambda+i\kappa}$ (recall that $\kappa$ is either $0$ or $\pi/\tau$). Therefore, in the case of an impurity, the equation for $\alpha$ reads
\begin{equation}
  -t_{x_0+2,x_0+1}(\xi,\tau_1)\frac{\psi(x_0+2)}{\psi(x_0+1)}+V_{x_0+1}+U=\epsilon(i\lambda+\kappa).\label{solvealphaU}
\end{equation}
For bulk modes, the same procedure is applied, with the eigenfunctions given instead by Eq.~(\ref{eq:LeftBulk}), and the quantisation condition for the quasi-momenta $k$ is given by
\begin{equation}
  -t_{x_0+2,x_0+1}(\xi,\tau_1)\frac{\psi_k(x_0+2)}{\psi_k(x_0+1)}+V_{x_0+1}+U=\epsilon(k),
\end{equation}
completely analogous to the edge modes.

Before tackling some examples, it is worth discussing the robustness of topological states. It is often, and correctly, stated that topological states are robust against weak local perturbations \cite{Hasan2010,Asboth2016,Bansil2016}. Of course, it is reasonable to expect that this assumption will break down for an infinite perturbation, i.e. a wall. However, the energy scale of the system is well defined by a finite bulk band gap \cite{Black-Schaffer2012}. It is straightforward to see that the topological edge state will be affected by the impurity we are considering. The final term on the left-hand side of Eq.~\eqref{solvealphaU} will lead to an energy contribution of the order of $U$ to the edge state. Of course, this in itself will not destroy the topological state, but simply change the energy of it within the bulk band gap. However, if the impurity is of the order of the energy difference between the original topological state and the bottom (repulsive $U$) or top (attractive $U$) of the next infinite-size bulk band, then the original topological state may now have an energy which lies in the bulk band. A state with an energy in the infinite-size bulk band has a real well-defined momentum, i.e. $\kappa=0$. Therefore, the original topological edge state can no longer be bound to the boundary and is hence no longer an edge (or topological) state.

\section{Su-Schrieffer-Heeger model}
In this section we investigate the Su-Schrieffer-Heeger (SSH) model \cite{Su1979}, a special limit of the Rice-Mele \cite{Rice1982} and Harper \cite{Harper1955} models. This was recently realized in an ultracold atom experiment by the group of Gadway, including the probing of the edge mode of the model \cite{Meier2016}. The effect of an impurity at an edge on the SSH model energy spectrum has been previously considered numerically \cite{Liberto2016}. The SSH model is very instructive since it is the simplest model that supports non-trivial topology. The off-diagonal Harper model has a Hamiltonian of
\begin{align}
H = \sum_x \left( t_{x,x+1} \hat{c}^{\dagger}_x \hat{c}_{x+1} + \mathrm{H.c.} \right)
\end{align}
where we will take $x=0,1,\dots,L_s-1$ for concreteness ($x_0=-1$) and where
\begin{align}
t_{x,x+1} = t + \Lambda \cos\left(\frac{2 \pi x}{\tau} + \theta \right),
\end{align}
with $\Lambda$ being the amplitude of the periodic modulation of the tunneling and $\theta$ a constant phase. For the SSH model we take $\tau = 2$ and $\theta = \pi$, and we will set $\lambda = \Lambda / t$ throughout. This results in 
\begin{align}
t_{x,x+1} = 1 - \lambda \left(-1\right)^x
\end{align}
and the stationary Schr\"odinger equation of
\begin{align}
\left[1 - \lambda \left(-1\right)^{x-1} \right] \psi\left(x-1\right) + \left[x - \lambda \left(-1\right)^{x} \right] \psi\left(x+1\right) \nonumber \\ = \varepsilon \psi\left(x\right),\label{eq:SSHSchrodinger}
\end{align}
with $\varepsilon \equiv E / t$. It is known that the SSH model is topological if the inter-site tunnelling is larger than the intra-site tunnelling \cite{Shen2012,Li2014,Yan2014,Liberto2016,Asboth2016,Chiu2016}. This has been confirmed by a direct measurement of the Zak phase \cite{Atala2013} and an observation of the topological state \cite{Meier2016} in cold atoms. In our notation, the SSH model is topological if $\lambda>0$, which we will set throughout. The topological invariant of this model is the winding number and it is zero for $\lambda<0$ and one for $\lambda>0$.

We begin by solving the Schr{\"o}dinger equation in the infinite size limit. Using Bloch's theorem, we write the eigenfunctions as $\psi_{k}(x)=\phi_k(x)\exp(ikx)$ with $\psi_{k}(x+2)=\psi_k(x)$. Inserting this into the Schr{\"o}dinger equation, we obtain, after minor algebraic manipulations,
\begin{align}
\phi_{s,\pm k} \left(x\right) = \begin{pmatrix}
1 \\
\frac{\left( 1 - \lambda\right) e^{\mp ik} + \left( 1 + \lambda\right) e^{\pm i k}}{\varepsilon_s}
\end{pmatrix},\label{eq:phiSSH}
\end{align}
with $s=0,1$ labelling the two bands of the system. For the energy bands we obtain
\begin{align}
\varepsilon = (-1)^s \sqrt{\left(2\cos k\right)^2+\left(2\lambda\sin k\right)^2}.
\label{eq:SSHEnergy}
\end{align}
The bulk states in the finite case, i.e. $x_0=-1$ and $L_s<\infty$ are then given by Eq.~(\ref{eq:LeftBulk}), with the Bloch functions in Eq.~(\ref{eq:phiSSH}), while the quasi-momentum is quantised according to Eq.~(\ref{eq:BulkQuantisation}).

\subsection{Solving for the States}
\label{sec:SSHImpurity}

As already discussed the edge states of the system can be constructed analytically, with the general form of Eq.~(\ref{eq:BoundState}). If we consider the case of a fully homogeneous lattice ($\lambda \equiv 0$), the conformal transformation
\begin{align}
\varepsilon = - 2 \cos k = - \frac{1+\alpha^2}{\alpha},
\label{eq:FlatEnergy}
\end{align}
maps the range $0 < \mid \alpha \mid < 1$ into $-2 \leq \varepsilon \leq 2$ \cite{Manuel2010}. We can utilise this mapping to transform the bulk energies with $k$ dependency to that of the $\alpha$ dependent edge states.

Substituting the expression for $\cos k$ of Eq.~(\ref{eq:FlatEnergy}) into Eq.~(\ref{eq:SSHEnergy}), we obtain the following form for the energy of the bound states
\begin{align}
\varepsilon_s = (-1)^s \sqrt{\left( \frac{1+\alpha^2}{\alpha}\right)^2 - \left(\lambda \frac{\alpha^2-1}{\alpha}\right)^2},
\label{eq:EnergyFunctional}
\end{align}
which agrees with the derived form starting with an ansatz of Eq.~\eqref{eq:SingleAnsatz}. For the case of no impurity ($U=0$), $\alpha$ is extracted by solving Eq.~(\ref{eq:U0Solve}) with the energy given by Eq.~(\ref{eq:EnergyFunctional}), while for $U\ne 0$ we must use Eq.~(\ref{solvealphaU}). Before moving on we need to obtain the Bloch functions $\phi_{\pm}$ for the bound states. To do this we substitute the infinite-size limit for the edge states, Eq.~(\ref{eq:SingleAnsatz}), into the Schr\"{o}dinger equation (\ref{eq:SSHSchrodinger}) with periodic boundary conditions ($\tau=2$) for $\phi_{\pm}$. We obtain
\begin{align}
\phi_{s,+} \left(x\right) = \begin{pmatrix}
1 \\
\frac{\left(1+\alpha^2\right) - \lambda \left(1-\alpha^2\right)}{\varepsilon_s \alpha}
\end{pmatrix},\label{sshsomeequation}
\end{align}
with $|\alpha|<1$, while for $\phi_{s,-}$ we carry out the transformation $\alpha \rightarrow 1/\alpha$ on Eq.~(\ref{sshsomeequation}).

\subsection{The Effect of an Impurity}
Due to the simplicity of this model, it is instructive to treat the case of an impurity of strength $U$ at $x_0+1$ explicitly, even though this has been done in general in the previous sections. We will take the impurity to be placed at $x=0$, with the boundary conditions $\psi\left(-1\right)=0$ and $\psi\left(L_s\right)=0$. The Schr\"odinger equation of the SSH model with an impurity at $x_0+1$ is given by
\begin{align}
\left[1 - \lambda \left(-1\right)^{x-1} \right] \psi\left(x-1\right) + \left[1 - \lambda \left(-1\right)^{x} \right] \psi\left(x+1\right) \nonumber \\ + \frac{U}{t} \delta_{x,x_0+1} \psi \left(x\right) = \varepsilon_s \psi\left(x\right).\label{eq:SSHSchrodImpurity}
\end{align}
We solve Eq.~(\ref{eq:SSHSchrodImpurity}) for the energy at $x=0$, obtaining 
\begin{align}
\varepsilon_s = \frac{U}{t} + \left(1- \lambda \right) \frac{\psi \left(1\right)}{\psi \left(0\right)}.\label{eq:SSHEnergyStates}
\end{align}
For the edge states we substitute Eq.(\ref{eq:LeftEdge}) into Eq.~(\ref{eq:SSHEnergyStates}), taking the energy to be given by Eq.~(\ref{eq:EnergyFunctional}) and solve for $\alpha$. Similarly, for the bulk states we substitute Eq.(\ref{eq:LeftBulk}) into Eq.~(\ref{eq:SSHEnergyStates}), taking the energy to be given by Eq.~(\ref{eq:SSHEnergy}) and solve for $k$.

\begin{figure}[t]
\begin{center}
\includegraphics[width=0.49\textwidth]{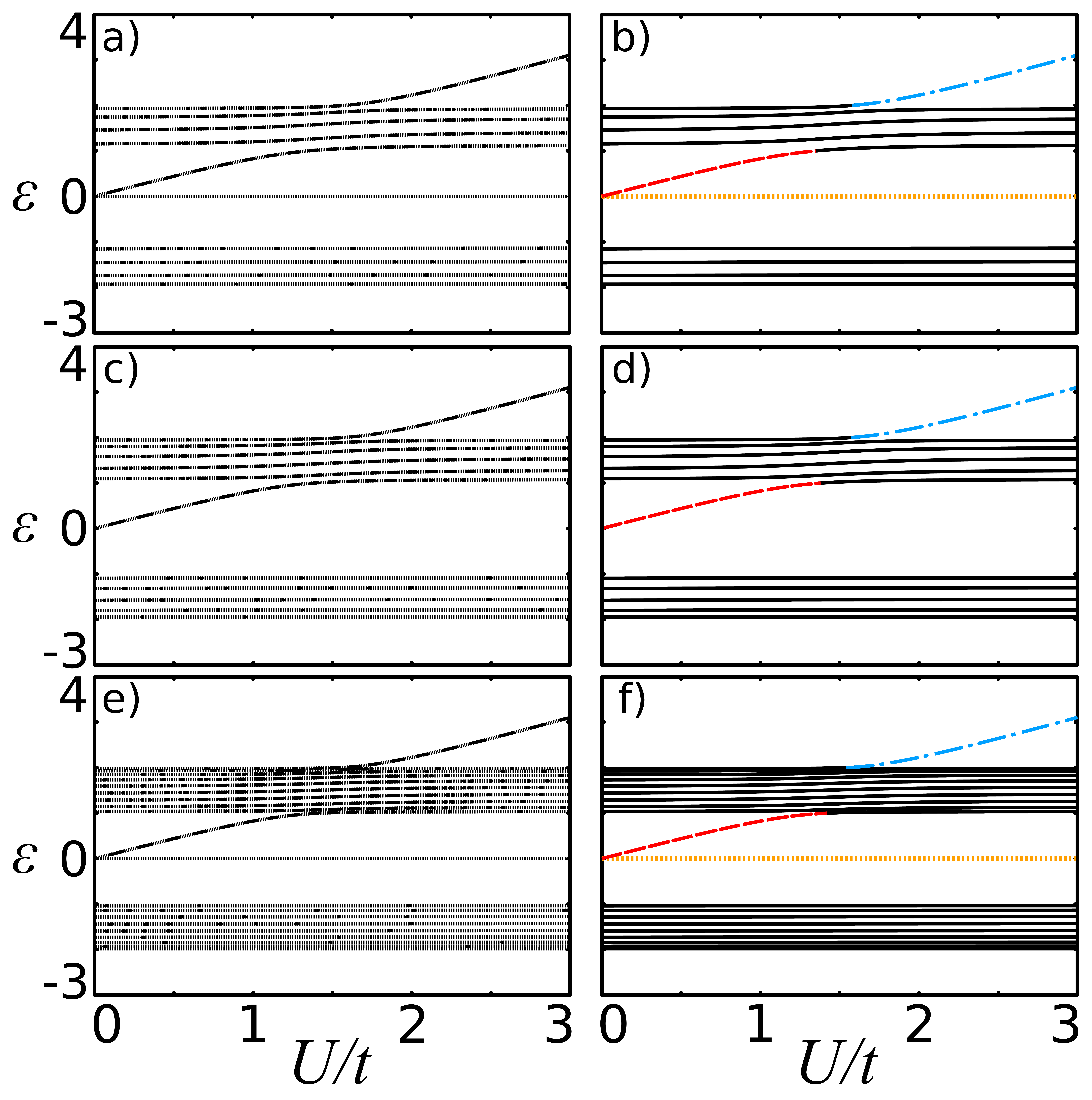}
\end{center}
\caption{Energy spectrum of the SSH model with an impurity $U$. Parameters are $\lambda = 0.5$ and a,b) $L_s=10$ c,d) $L_s=11$ e,f) $L_s=20$. a,c,e) Exact diagonalization of the Hamiltonian numerically. b,d,f) The analytical approach, with bulks states solid (black) lines, the original edge $U=0$ left and right edge state with phase $\pi/\tau$ given by dashed (red) and dotted (orange) lines and the Shockley state with $0$ phase by a dashed-dotted (blue) line.}
\label{fig:SSHEnergies}
\end{figure}

\begin{figure}[t]
\begin{center}
\includegraphics[width=0.49\textwidth]{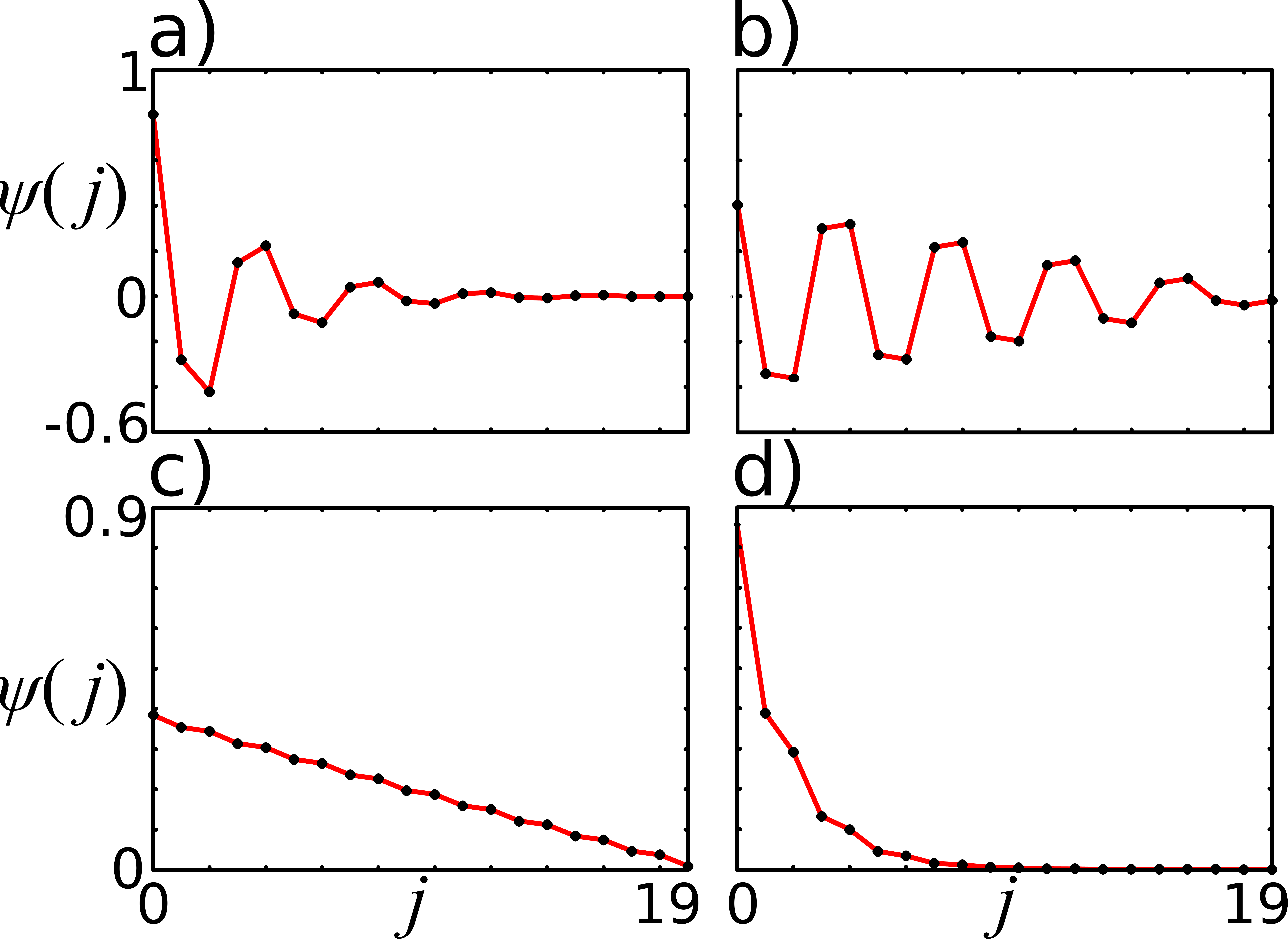}
\end{center}
\caption{Wave functions of the edge/bound state at $j=0$ of the SSH model, with $\lambda = 0.5$ and $L_s=20$. a) $U=1$ with $\alpha = 0.725 e^{i \frac{\pi}{2}}$, b) $U = 1.42$ with $\alpha = 0.977 e^{i \frac{\pi}{2}}$, c) $U=1.54$ with $\alpha = 0.978$ and d) $U=2$ with $\alpha = 0.583$. The solid (red) lines are a guide for the eye between the analytical points and the circles (black) are from solving numerically.}
\label{fig:SSHStates}
\end{figure}

We observe a perfect agreement between the method proposed in this work and the results of numerical exact diagonalisation of the system. This is shown for the spectra in Fig.~\ref{fig:SSHEnergies} and the topological edge and Shockley states in Fig.~\ref{fig:SSHStates}.

First, we consider the behaviour of the edge states. As the impurity strength is increased, the topological edge state bound near $x=0$, whose energy lies in the gap, increases in energy. This is observed for any system size, as seen in Fig.~\ref{fig:SSHEnergies}, including for $L_s$ that is incommensurate with the periodicity $\tau$ ($L_s\ne 2n -1$, $n\in \mathbb{Z}_{+}$). We observe that even a weak impurity  has an effect on the system but as expected does not destroy the topological state. The $L_s=\tau n$ system, in this case, are two edge states that for $U=0$ are hybridised into a symmetric and antisymmetric pair \cite{Asboth2016,Rhim2017a}. The introduction of even a small impurity ($U \sim 0.01$) breaks the symmetry that leads to this hybridisation, splitting the edge states into two states localised separately to the $x=0$ and $x=L_s-1$ edges (left and right) of the system respectively. In the SSH model, there is little effect of the impurity on the state now localised to the right edge, due to its negligible probability density at the location of the impurity.

With increasing $U$ the edge state bound to the $x=0$ edge eventually merges with the bulk resonantly, i.e. $|\alpha|=1$ at the merging point. The resonant value of $U/t$ is inferred from Fig.~\ref{fig:SSHEnergies}, and corresponds to the value at which the in-gap red-dashed line (edge mode) disappears. As expected, the resonant value of the impurity strength is of the order of the bulk band gap. In the Bloch state picture, the edge state resonantly disappears when its quasi-momentum goes from being complex to real, i.e. the wave function goes from one that is localised (`decaying') to one that is delocalised. This transition is indeed observed in the form of the states in Fig.~\ref{fig:SSHStates}a and b.

As the edge state is merging with the bulk band, the ejection of a state from the top of the band takes place. This state is trivially bound to the impurity, hence, it is a Shockley state \cite{Shockley1939}. This is most easily seen in the limit of large $U$, in which $\psi\left(0\right)\rightarrow 1$ while the states eigenenergy is $\varepsilon\sim U/t$. This state is distinctive as it corresponds to a real value of $\alpha$ ($\kappa=0$). As $U$ is increased, as expected, we observe the state becoming further localised to the left edge, as seen in Fig.~\ref{fig:SSHStates}c and d.

It comes as no surprise that the bulk states are changed in a more subtle way by the impurity. Away from regions of merging and ejection of bound states, little changes in their energy. However, around the point where the left edge state merges with the bulk, the energies of the bulk states are pushed up to accommodate the addition of the state from below and the ejection of a state from the top of the band. This effect is clearly observed in Fig.~\ref{fig:SSHEnergies}b and d.

\section{Diagonal Harper Model}

We now move on to consider the diagonal Harper, or Aubry-And\'{e}, model consisting of a periodic modulation of the on-site potential and a constant tunneling rate. This model corresponds to the tight-binding approximation of a particle in a superlattice, and has been realised in photonic lattices \cite{Lahini2009,Kraus2012} and trapped ultracold atomic systems \cite{Schreiber2015}. The dimensionless Hamiltonian of the system is given by
\begin{align}
H = - \sum_x \left( \hat{c}^{\dagger}_x \hat{c}_{x+1} + \mathrm{H.c.} \right) + \sum_x \frac{V_x\left(\Lambda,\tau,\theta\right)}{t} \hat{c}^{\dagger}_x \hat{c}_{x},
\label{eq:DiagHarperH}
\end{align}
with $x=0,\dots,L_s-1$ (i.e. we set $x_0=-1$ for concreteness), and the periodic potential given by
\begin{align}
V_x\left(\Lambda,\tau,\theta\right) = \Lambda \cos\left(\frac{2 \pi x}{\tau} + \theta\right).
\end{align}
Above, $\theta$ is a constant phase and $\tau$ the spatial periodicity. We will consider the dimensionless potential strength $\lambda = \Lambda/t$ throughout. It is known that the 1D Harper model is a single phase component of the 2D Hofstadter model and that the topological nature of the Hofstadter model is captured by this model \cite{Mei2012,Goldman2015}.

\subsection{The Effect of an Impurity}

\begin{figure}[t]
\begin{center}
\includegraphics[width=0.49\textwidth]{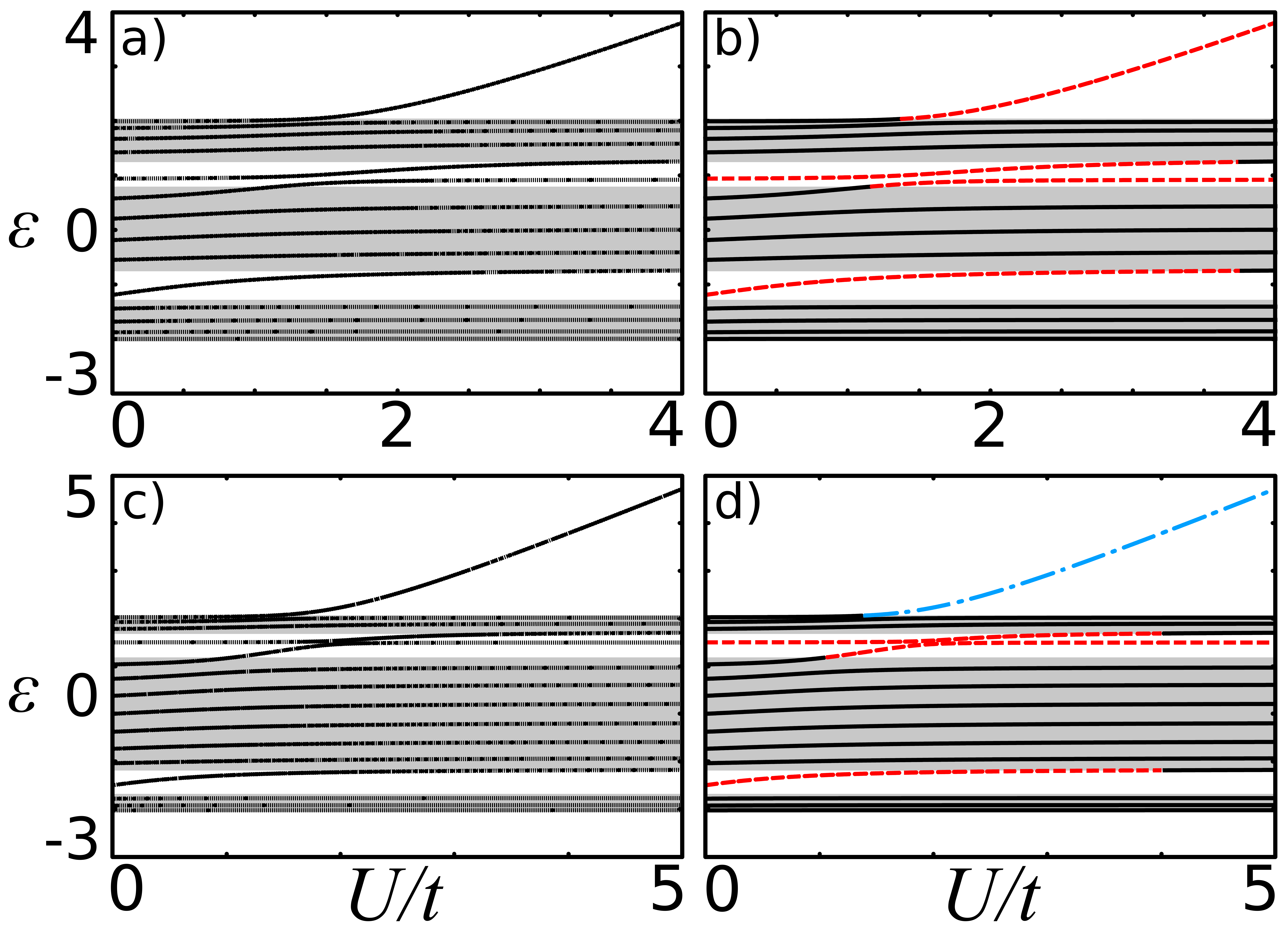}
\end{center}
\caption{Energy spectrum of the Harper model with an impurity, $U$, shaded areas represent the continuum bulk bands. Values of $\lambda = 0.5$, $\theta = \pi$ and a,b) $\tau=3$ $L_s=14$ c,d) $\tau=4$ $L_s=15$. a,c) Numerical exact diagonalization. b,d) Analytical approach with bulks states solid (black) lines, edge states of phase $0$ with dashed-dot (blue) lines and edge states of phase $\pi/\tau$ with dashed (red) lines.}
\label{fig:HarperEnergies}
\end{figure}

We consider the effect of an impurity at $x=0$ in the Harper model. In the following examples we study the cases of $\tau=3$ and $\tau=4$, both with $\lambda=1/2$ and $\theta = \pi$. We choose $L_s=14$ and $L_s=15$ for $\tau=3$ and $\tau=4$, respectively. These sizes correspond to the commensurate case in the limit $U=0$, while they become effectively incommensurate ($L_s\to L_s-1$) in the limit $U\to\infty$ after dropping the trivial Shockley state that is ejected from the highest band at finite $U/t$ (see Fig.~\ref{fig:HarperEnergies}).

The behaviour of the spectra as the impurity strength is increased is shown in Fig~\ref{fig:HarperEnergies}. There, numerical exact diagonalisation results are plotted on the left panels for comparison (Fig.~\ref{fig:HarperEnergies}a and c), while the results using our method are plotted on the right panel (Fig.~\ref{fig:HarperEnergies}b and d). Again, we observe a perfect agreement between numerics and the detailed method of this work. The energies of bound states corresponding to real values of $\alpha$ are plotted in dash-dot (blue) lines, while for $\alpha=|\alpha|\exp(i\pi/\tau)$ their energies are plotted in dashed (red) lines. For $\tau=3$ (Fig.~\ref{fig:HarperEnergies}a and b), at $U=0$ there are three bands (shaded regions) where the bulk modes lie, and two edge modes, one per band gap. As the impurity strength is increased, the edge mode in the first band gap has a corresponding energy increase until it disappears into the bulk, at $U/t=3.76$. The behaviour of the energy of the edge mode in the second band gap is quite different for $0<U/t<1$, as it is essentially unchanged by the impurity, indicating that in this regime this edge mode is localised to the opposite edge. At $U/t=1.15$, however, a further bound state appears in the second band gap, pushing the other state upwards in energy, where eventually it merges with the bulk at $U/t=3.76$, the same value as the other original edge state. The state appearing at $U/t=1.15$ approaches, asymptotically ($U/t\to \infty$) the energy of an edge state in the commensurate case, indicating that it is bound to the $L_s-1$ (right) edge of the system. These two levels exhibit an avoided crossing in which their characters -- left- and right-edge localisation -- are eventually interchanged. This effect may be interesting from a Landau-Zener perspective. At $U/t=1.36$, a Shockley state appears above the highest band of the system, and its energy approaches, asymptotically, $\varepsilon\sim U/t$. 

\begin{figure}[t]
\begin{center}
\includegraphics[width=0.49\textwidth]{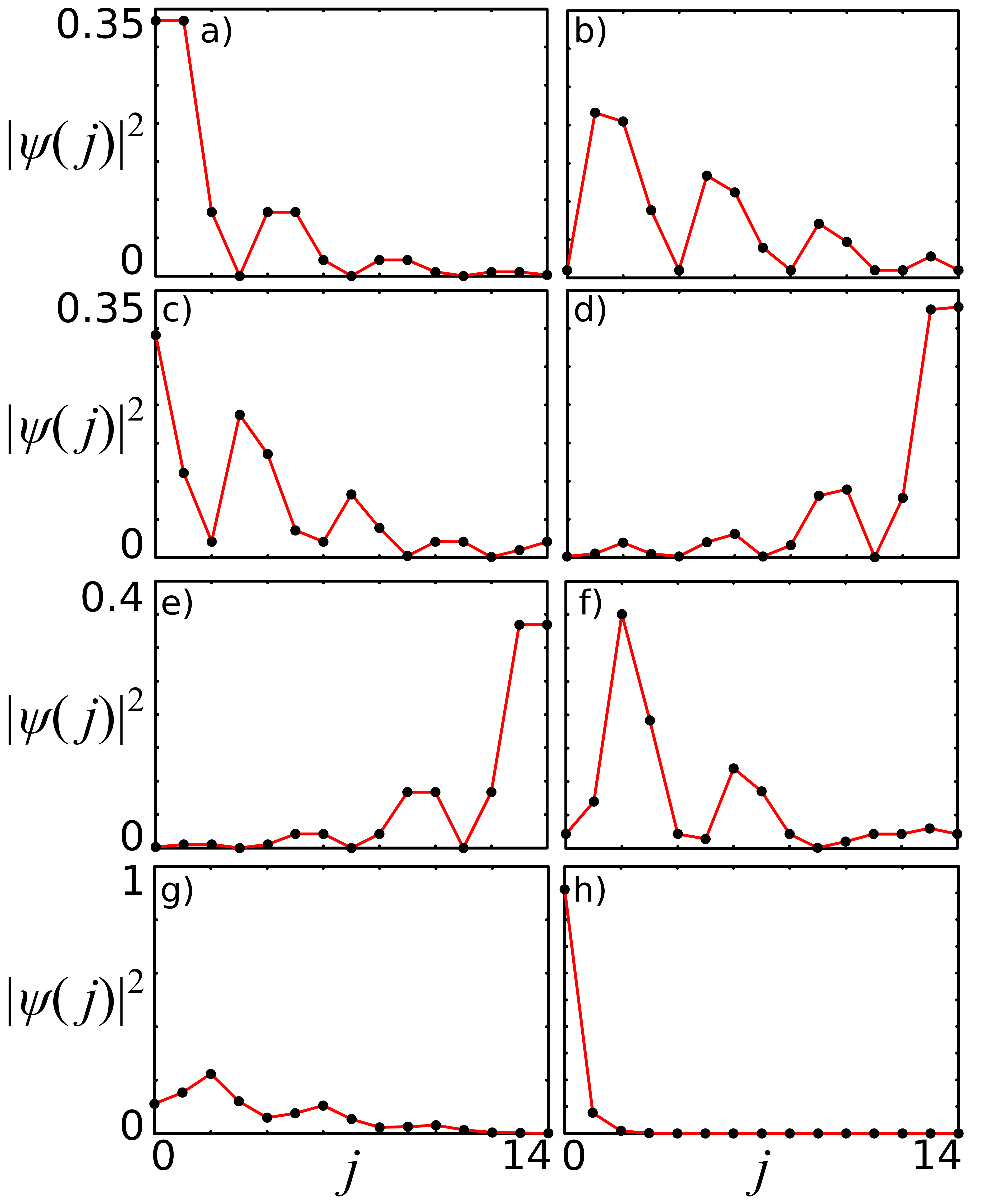}
\end{center}
\caption{The edge states of the diagonal Harper model for $\tau=4$, $L_s = 15$ with impurity $U$. a,b) $U=0$ edge state in the first band gap for a) $U=0$ and b) $U=4$, c,d) emerging state in second band gap that emerges from the band for c) $U=1.07$ and d) $U=4$, e,f) $U=0$ edge state in the second band gap at e) $U=0$ and f) $U=4$ and g,h) Shockley state emerging from second band for g) $U=1.39$ and h) $U=4$. Each state has $\varepsilon$ and $\alpha$ of; a) $\varepsilon=-1.500$, $\alpha = 0.841 e^{i \frac{\pi}{4}}$, b) $\varepsilon=-1.186$, $\alpha = 0.999999 e^{i \frac{\pi}{4}}$, c) $\varepsilon=1.187$, $\alpha = 0.990 e^{i \frac{\pi}{4}}$, d) $\varepsilon=1.495$, $\alpha = 0.840 e^{i \frac{\pi}{4}}$, e) $\varepsilon=1.500$, $\alpha = 0.841 e^{i \frac{\pi}{4}}$, f) $\varepsilon=1.686$, $\alpha = 0.999999 e^{i \frac{\pi}{4}}$, g) $\varepsilon=2.062$, $\alpha = 0.987$, h) $\varepsilon=3.789$, $\alpha = 0.287$.}
\label{fig:HarperEdge}
\end{figure}

For $\tau=4$, Fig.~\ref{fig:HarperEnergies}c and d, there are four energy bands. However, two of them cross forming a single continuum (shaded area) in the centre of the spectrum. The behaviour of the spectrum as the impurity strength is cranked up is qualitatively identical to the case of $\tau=3$, with the original edge states merging with the bulk bands at $U/t=4.01$, a bound state emerging in the second band gap at $U/t = 1.068$ and a Shockley state emerging from the highest band at $U/t=1.388$. We now move on to discuss the form of the edge states, whose probability densities are plotted in Fig.~\ref{fig:HarperEdge}. In Fig.~\ref{fig:HarperEdge}a and Fig.~\ref{fig:HarperEdge}e, we plot the probability densities for the topological edge modes in the first and second gap, respectively, at $U/t=0$. As was discussed through their energetic properties, these states are bound near the left and right edge of the system, respectively, when the impurity is absent, a fact that is clearly observed in the figures. In Fig.~\ref{fig:HarperEdge}b we show the edge state in the first band gap for $U/t=4$, a value very close to the resonance, and hence its large extension inside the bulk of the system. The probability density of the state that emerges from the middle bulk bands into the second band gap at $U/t=1.068$ is plotted for near-resonant coupling ($U/t=1.07$) in Fig.~\ref{fig:HarperEdge}c where, again, we see that it is rather extended. In Fig.~\ref{fig:HarperEdge}d we show this state in the strong-coupling limit ($U/t=4$), where it clearly becomes localised to the right edge with a probability density that is perturbatively close to that in Fig.~\ref{fig:HarperEdge}e, discussed above. The opposite effect happens to the edge mode that exists for $U/t=0$ in the second band gap. For $U/t=4$ (near the resonance), this edge mode has moved towards the left, and its extension has become larger. The probability density of the trivial Shockley state is plotted in Fig.~\ref{fig:HarperEdge}g and h, where we observe increasing localisation to $x=0$ as the impurity strength goes up, as expected. 

\section{Two-Dimensional Models}

\begin{figure}[t]
\begin{center}
\includegraphics[width=0.49\textwidth]{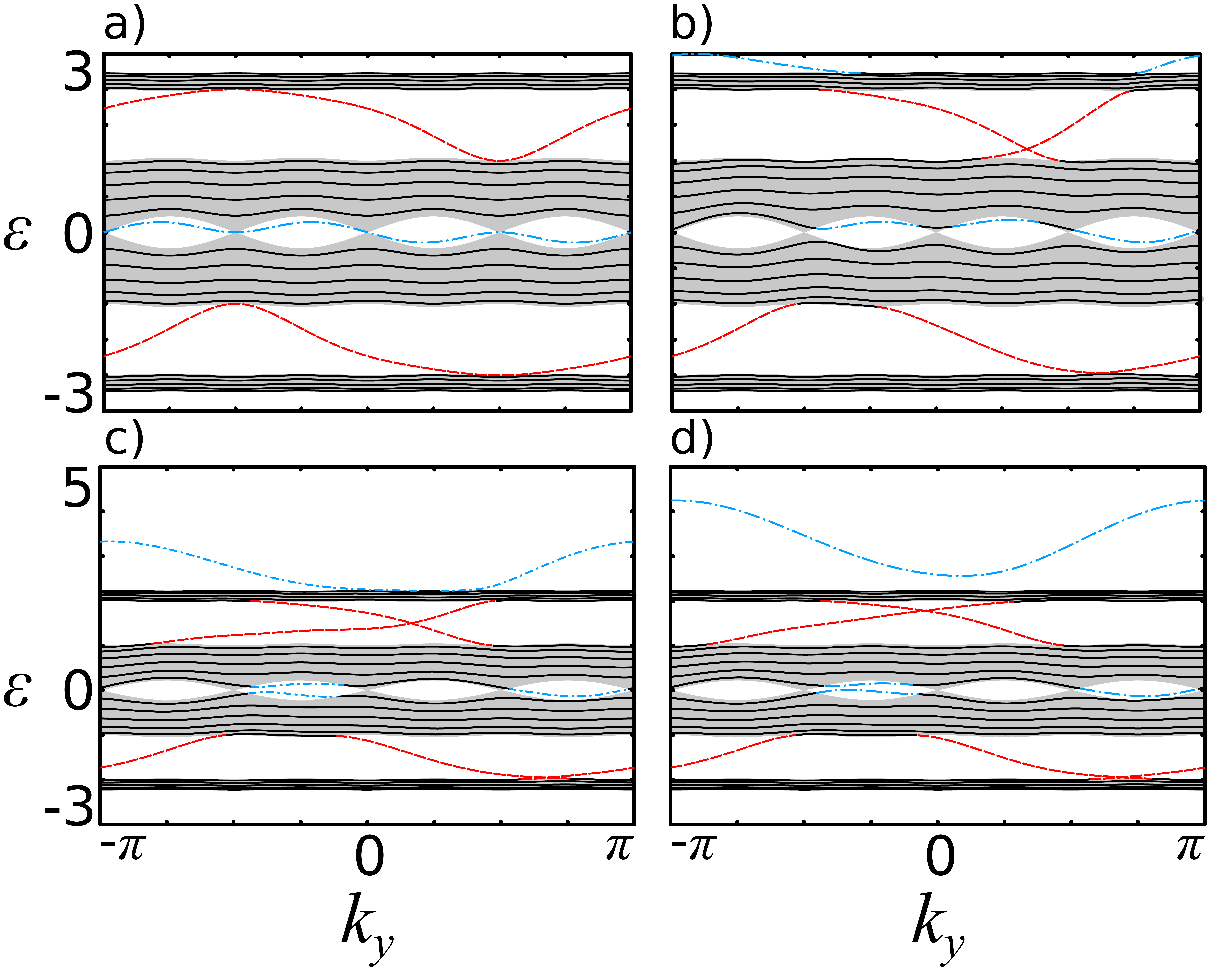}
\end{center}
\caption{Spectra for the Hofstadter model obtained via the exact method for $t = 1$, $\lambda = 0.5$, $L_s=23$ and an impurity of a) $U=0$, b) $U=1$, c) $U=2$, d) $U=3$. Bulk states given by solid (black) lines, edge states with phase $\pi/\tau$ by dashed (red) lines and edge states with phase $0$ by dashed-dotted (blue) lines.}
\label{fig:BandsHofstadter}
\end{figure}

We consider next two-dimensional (2D) periodic models on a square lattice described by Hamiltonians of the form
\begin{align}
H = - \sum_{x,y} \left[ t_1 \hat{c}_{x+1,y}^{\dagger} \hat{c}_{x,y} + t_2\left( x\right) \hat{c}_{x,y+1}^{\dagger} \hat{c}_{x,y} + \mathrm{H.c.} \right], \label{eq:TwoDHamiltonian}
\end{align}
where we have set the lattice spacing $d=1$, with $t_1 = t$ a constant and $t_2\left(x+\tau\right) = t_2\left(x\right)$.  $\hat{c}_{x,y}$ ($ \hat{c}_{x,y}^{\dagger}$) annihilates (creates) a particle at site $\left(x,y\right)$. We particularise to the Hofstadter model \cite{Hofstadter1976} in the Landau gauge, for which the periodic tunnelling function $t_2$ can be written as
\begin{align}
t_2 \left( x \right) = \Lambda e^{i \Omega\left(x\right)},
\end{align}
where $\Lambda$ is a constant and $\Omega\left(x\right) = 2\pi x p/q$ with $p$, $q \in \mathbb{Z}$. As before, we will study the properties of the model in terms of the dimenionless parameter $\lambda=\Lambda/t$. The Hofstadter model has been realised in ultracold atom experiments by the group of Ketterle \cite{Miyake2013,Kennedy2015}. This model is well known to be topological \cite{Hofstadter1976,Thouless1982,Hafezi2007,Hugel2014}, with each band of the phase diagram having a nonzero Chern number \cite{Goldman2015}. The non-zero Chern numbers of the bands in the Hofstadter model have been directly measured in cold atoms \cite{Aidelsburger2015}.

We take the boundary conditions to be those of an infinite strip geometry, with the system being boundary-free in the $y$-coordinate and with sharp edges in the $x$-coordinate. In this geometry, the stationary Scr\"odinger equation $H\Psi=E\Psi$ for the 2D Hamiltonian of Eq.~(\ref{eq:TwoDHamiltonian}) can be separated by writing
\begin{align}
\Psi \left(x, y\right) = \psi \left(x\right) e^{i k_y y},\label{psihof}
\end{align}
with $k_y$ being a quasi-momentum with $k_y \in \left[ 0,2\pi\right)$. Substituting Eq.~(\ref{psihof}) into the stationary Schr{\"o}dinger equation for Hamiltonian (\ref{eq:TwoDHamiltonian}), we obtain a one dimensional Schr\"odinger equation corresponding to a parametric (in $k_y$) diagonal Harper model, Eq.~(\ref{eq:DiagHarperH}), with the on-site periodic potential $V_x$ given by
\begin{align}
V_x = -2 \lambda \cos \left( \Omega\left(x\right) - k_y \right).
\label{eq:HofstadterPotential}
\end{align}
The edge and bulk states of the Hofstadter model is equivalent to those of the diagonal Harper model, albeit with a phase $\theta=-k_y$. The case of an impurity next to the edge in Harper's model becomes now that of a line defect. This is very relevant in this system, as it may be viewed as the minimal model for a softened edge potential. As we have already discussed the Harper model in some detail we will move straight on to considering the effect of an impurity in the Hofstadter model.

\subsection{The Effect of an Impurity}

We will consider as an example the case of $x_0=-1$, $L_s=23$ (commensurate for $U/t=0$), $\lambda=1/2$ and $\tau = 4$ (or $p/q=1/4$). Spectra corresponding to four different values of the impurity strength $U/t$ are shown in Fig.~\ref{fig:BandsHofstadter}.

From the results of the diagonal Harper model, examples of which were presented in the previous section for $k_y=-\pi$ ($\theta=\pi$), we would expect the edge states localised around $x=0$ to survive the presence of an impurity in the Hofstadter model only for some quasi-momenta $k_y$, while those near the right end of the system are expected not to be affected in any visible way. This is indeed the case. In Fig.~\ref{fig:BandsHofstadter} ($U/t=0$), we have one edge mode within each of the band gaps, as is known, since the edge mode spectrum coincides, for the commensurate case, with that on a semi-infinite ($L_s=\infty$) plane \cite{Hugel2014}. As observed in Fig.~\ref{fig:BandsHofstadter}b-d, the impurity changes the spectrum of bound states qualitatively. For small-to-moderate impurity strength ($U/t=1$, Fig.~\ref{fig:BandsHofstadter}b), there are different regions of quasi-momenta where edge states disappear and none are remaining, in each of the band gaps, while two separate branches of edge states are now present in the highest energy gap. A band of Shockley states appears above the highest continuum band of the system. This phenomenology is also present for stronger coupling $U/t$ (Fig.~\ref{fig:BandsHofstadter}c-d), with the addition of further edge modes in the first and second band gaps in some regions of quasi-momentum. Note, that at points where the edge states cross in Fig.~\ref{fig:BandsHofstadter}b, c and d, there is the opening of a small gap of the order of $\epsilon \sim 0.001$ or smaller. The size of this gap decreases with increasing impurity strength and system size.

\section{Transition probabilities after a quench}

In order to gain further understanding of the robustness of edge modes or absence thereof, it is important to discuss feasible ways of probing their stability when an impurity or a potential near the edge is present. In particular, the transition or decay probability of an initially prepared edge mode into the bulk continua when the system is perturbed is a relevant measure of robustness. Moreover, its calculation shows the power of knowing the exact analytical form of all eigenfunctions for simplifying related problems, especially for more than one particle and in higher dimensions. For concreteness and simplicity, we focus on the diagonal Harper model in 1D. Extension to other 1D and reducible 2D models are straightforward.  

We consider a model that has an initial Hamiltonian of Eq.~(\ref{eq:DiagHarperH}) with the addition of a perturbation that starts at time $T=T_1$. The perturbing operator corresponds to a potential acting on the lattice well next to the left ($x_0+1$) edge of the system. The Hamiltonian of the system is then given by
\begin{align}
H^{\prime}\left(T\right) =H+ \frac{\mathcal{U}}{t} \theta(T-T_1) \hat{c}^{\dagger}_{x_0+1} \hat{c}_{x_0+1},\label{eq:Pert}
\end{align}
where $H$ is the diagonal Harper's Hamiltonian in dimensionless form, Eq.~\eqref{eq:DiagHarperH}, $\mathcal{U}$ is an impurity strength constant, $T$ denotes time and $T_1$ is the time when the perturbation is switched on, while $\theta(T-T_1)$ is the Heaviside step function. 

The transition probability for an initial state $\ket{i}$, which is an eigenstate of the Hamiltonian for $T<T_1$, to transition into a set of final states $F=\{\ket{f_{\mathcal{U}}}\}$, all of which are eigenstates for $T>T_1$, can be written as
\begin{align}
P_{i\rightarrow F}\left(T\right) =\sum_{f_{\mathcal{U}}\in F} \lvert \braket{f_{\mathcal{U}}} \ket{i(T)} \rvert^2,
\label{eq:TransitionProb}
\end{align}
where $\ket{i(T)}$ satisfies the time-dependent Schr{\"o}dinger equation at time $T>T_1$ with initial condition $\ket{i(T_1)}=\ket{i}$. We notice that, since the spectrum of edge and bulk modes is not only quantitatively but also qualitatively changed by the presence of a static potential, the overlaps in Eq.~(\ref{eq:TransitionProb}) must be between the time-evolved initial state and a set of final states corresponding to eigenstates of the quenched Hamiltonian. The transition probability in Eq.~(\ref{eq:TransitionProb}) is time-independent, which is seen by writing $\ket{i(T)}=\hat{U}(T,T_1)\ket{i}$, where $\hat{U}(T,T_1)$ is the propagator for Hamiltonian (\ref{eq:Pert}) for $T>T_1$. Since $\ket{f_{\mathcal{U}}}$ are eigenstates of the perturbed Hamiltonian, the transition probability takes the form
\begin{align}
P_{i\rightarrow F}\left(T\right) =\sum_{f_{\mathcal{U}}\in F} \lvert \braket{f_{\mathcal{U}}} \ket{i} \rvert^2,
\label{eq:TransitionProb2}
\end{align}
The most interesting process upon quenching the system is the transition of an initial topological edge state, $\ket{i}=\ket{\psi_e\left(T<T_1\right)}$ into the bulk. The transition probability is given by Eq.~(\ref{eq:TransitionProb2}) with the set of final states $F=\{\ket{f_{\mathcal{U}}}\}$ with $\ket{f_{\mathcal{U}}}$ eigenstates of the perturbed Hamiltonian with energies lying in the infinite-size limit continua (the bulk, by definition). We note that as we sum over all bulk states at time $T$, the transition probability from edge to bulk is not a continuous function of $\mathcal{U}/t$, as we have seen that the number of bulk states is not conserved as the perturbation of Eq.~(\ref{eq:Pert}) crosses a resonance. In fact, we shall see that the expected discontinuities constitute unambiguous measures of transition points where states enter or leave the bulk.

The calculations of the transition probabilities can be simplified using the analytical form of the eigenfunctions both in the perturbed and unperturbed situations. We shall consider, without loss of generality, $x_0=-1$ for the sake of clarity. For instance, the amplitudes $\braket{\psi_k^{\mathcal{U}}}\ket{\psi_{\alpha}}$, where $\psi_k^{\mathcal{U}}$ is a perturbed bulk state and $\psi_{\alpha}$ is an unperturbed edge mode, are given by
\begin{align}
  \braket{\psi_k^{\mathcal{U}}}\ket{\psi_{\alpha}}=& \sum_{x=0}^{L_s-1}\left[A_k^*\phi_k^*(x)e^{-ikx}+B_k^*\phi_{-k}^*(x)e^{ikx}\right] \nonumber \\ & \times \left[\bar{A}_{\alpha}\phi_+(x)\alpha^x+\bar{B}_{\alpha}\phi_{-}(x)\alpha^{-x}\right],\label{overlap}
\end{align}
where the constants $A_k$, $B_k$, $\bar{A}_{\alpha}$ and $\bar{B}_{\alpha}$ are related in such a way that the states are normalised and solutions to their respective stationary Schr{\"o}dinger equations, as discussed above at length. What is important here is the periodicity of $\phi_{\pm k}$ and $\phi_{\pm}$. By expanding Eq.~(\ref{overlap}) we obtain four different terms, all of which are proportional to sums of the form
\begin{equation}
  S=\sum_{x=0}^{L_s-1}\phi_1^*(x)\phi_2(x)e^{zx},\label{sumS}
\end{equation}
where the functions $\phi_1^*$ and $\phi_2$ have periodicity $\tau$. Given $L_s$, we find the largest positive integer $n$ such that $(n-1)\tau<L_s\le n\tau -1$. Using the periodicity property of $\phi_1^*$ and $\phi_2$, the sum $S$ in Eq.~(\ref{sumS}) can be rewritten as
\begin{align}
  S&=\sum_{j=0}^{\tau-1}\phi_1^*(j)\phi_2(j)e^{jz}\sum_{m=0}^{n-1-j}e^{mz\tau}\nonumber\\
  &+\sum_{x=(n-1)\tau+1}^{L_s-1}\phi_1^*(x)\phi_2(x)e^{zx}\nonumber\\
  &=\sum_{j=0}^{\tau-1}\phi_1^*(j)\phi_2(j)e^{jz}\frac{e^{z\tau(n-j)}-1}{e^{z\tau}-1}\nonumber\\
  &+\sum_{x=(n-1)\tau+1}^{L_s-1}\phi_1^*(x)\phi_2(x)e^{zx}.\label{sumSexplicit}
\end{align}
We see now that, by knowing the analytical form of the eigenfunctions and their properties we can simplify matrix elements involving them significantly. The basic sums in Eq.~(\ref{sumS}) have been simplified to sums of a very small number of terms, Eq.~(\ref{sumSexplicit}). Moreover, in the infinite-size case, edge-to-bulk matrix elements become even simpler to evaluate. In this case, the $\alpha^{-x}$ term ($|\alpha|<1$) for the edge states and the $e^{-ikx}$ term for the bulk states (Eqs.~(\ref{eq:BoundState}) and (\ref{eq:BulkState}), respectively) disappear. Notice that in this case we must normalise the bulk modes as $\braket{\psi_{s',k'}}\ket{\psi_{s,k}}=\delta_{s,s'}\delta(k-k')$, where we have reintroduced the band indices $s$ and $s'$. The sum in Eq.~(\ref{sumSexplicit}) then takes the form
\begin{equation}
S_{\infty}=\frac{1}{1-e^{z\tau}}\sum_{j=0}^{\tau-1}\phi_1^*(j)\phi_2(j)e^{jz}.
\end{equation}

\begin{figure}[t]
\begin{center}
\includegraphics[width=0.49\textwidth]{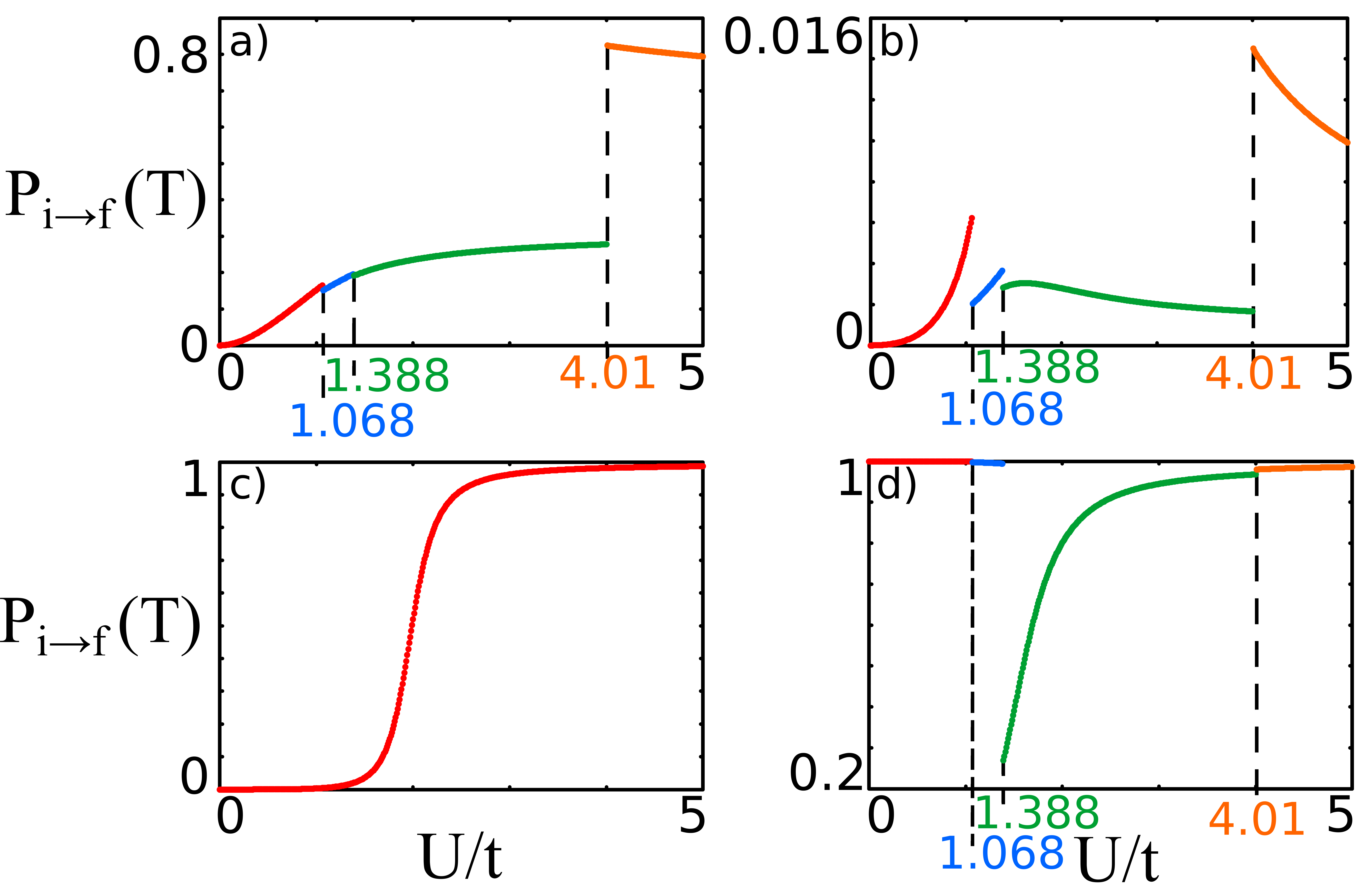}
\end{center}
\caption{Transition probabilities denoted by Eq.~(\ref{eq:TransitionProb}) for the perturbed diagonal Harper model with $L_s=15$ and $\tau=4$ from an initial state ($i$) at $T=0$ to a set of (or single) final states ($f$) at $T$. a) $i$ -- edge state with $\varepsilon=-1.5$, $f$ -- all bulk states of the perturbed Hamiltonian (denote as $B$), b) $i$ -- edge state with $\varepsilon=1.5$, $f$ -- $B$, c) $i$ -- edge state with $\varepsilon=1.5$, $f$ -- the 11th state, which starts in the bulk and in the $U \rightarrow \infty$ is in the band gap, see Fig.~\ref{fig:HarperEnergies}d, and d) $i$ -- highest energy state with $\varepsilon=2.03$, $f$ -- $B$. Points of discontinuity are marked on the plots and correspond to points where states enter/leave the bulk.}
\label{fig:OverlapsDiagHarper}
\end{figure}

We now consider a particular system, the diagonal Harper model. In order to connect with what was found in previous sections, we take $\tau=4$, $x_0=-1$ and $L_s=15$. These values correspond to the spectra in Fig.~\ref{fig:HarperEnergies}d and the probability densities of Fig.~\ref{fig:HarperEdge}. First we consider the transition probability of the lowest energy edge state (at $\varepsilon\approx -1.5$) for the unperturbed Hamiltonian (see Fig.~\ref{fig:HarperEnergies}d), to go into the bulk as a function of the quench strength. This is given by Eq.~(\ref{eq:TransitionProb2}) and shown in Fig.~\ref{fig:OverlapsDiagHarper}a. We observe the expected discontinuities in the probability when states enter or leave the bulk. For $\mathcal{U}/t>4$ we observe a large probability for the state transitioning to the bulk, while for $\mathcal{U}/t<4$ there is still non-negligible overlap with bulk states that would allow such a transition to the bulk. 

The behaviour of the transition probability as a function of the quench strength is very different for the second unperturbed edge state (at $\varepsilon\approx 1.5$, see Fig.~\ref{fig:HarperEnergies}d), as shown in Fig.~\ref{fig:OverlapsDiagHarper}b. Most strikingly the transition probabilities are of far lower magnitude than that of the lower energy edge state. Showing further that this edge state is stable under the perturbation. This comes as no surprise, as this edge state is localised to the wall away from the impurity (right edge), see Fig.~\ref{fig:HarperEdge}e. The low transition probability also matches up with the spectra in Fig.~\ref{fig:HarperEnergies}d, where the edge state is still present for large $U$. In fact, if we consider the overlap between the initial edge state and the state which moves into the second band gap with increasing potential strength, Fig.~\ref{fig:OverlapsDiagHarper}c, we observe an almost perfect transition of the initial edge state to the new band gap state.

We can also consider the properties of the Shockley state that appears out of the top of the highest band for $U \geq 1.4$ (see Fig.~\ref{fig:HarperEnergies}d). For this we consider the initial state to be the highest energy state of the spectrum -- a bulk mode of the unperturbed Hamiltonian -- and we calculate the overlap with all bulk states of the perturbed Hamiltonian. We observe that there is a significant probability of finding the state outside of the bulk for moderate $\mathcal{U}/t$, due to the Shockley state's energy being close to the energy at the top of the bulk. However, since the energy of the bound state that exits the band scales linearly with the impurity strength, a transition to this state becomes highly unfavourable for large $\mathcal{U}/t$. 

In this section, we have seen that, while for very weak impurity potentials edge states are robust, they can become unstable. Even for small-to-intermediate impurity strengths, the edge state can still have a non-negligible probability of going to the bulk under quenches of the form discussed here. It can also be seen that edge states away from the impurity are relatively stable. Their transition probabilities may exhibit large discontinuities at the critical potential strengths for which edge modes enter or exit the bulk. These ``critical points'' are nothing but resonances, and should have a major effect on the dynamics of these systems -- not only on the transition probabilities. As such, these resonances may be clearly identified in observables with the introduction of an impurity in topological models that have already been realised in cold atomic \cite{Meier2016,Schreiber2015,Miyake2013,Kennedy2015} and photonic lattice systems \cite{Lahini2009,Kraus2012}.

\section{Conclusions}

In this paper, we have established a method for obtaining the exact wave functions and spectrum of general one-dimensional periodic Hamiltonians and two-dimensional periodic models that are reducible to one dimension. The method described is very general and simple. It is also powerful in its ability to obtain all states and energies of finite periodic systems analytically, without any constraints or fine-tuning of its size, such as commensurability. For topologically non-trivial models of this kind, to which some of the most relevant systems belong, it is possible to extract topological states and their properties by their exact derivation. We show this in multiple examples of the SSH, diagonal Harper, and Hofstadter models. As the method is fully general, it is also possible to consider the effect of impurities on the system. In various examples of different topological models, we have shown that the symmetry-breaking placement of an impurity at an edge site of the lattice can have a drastic effect on the topological states of the system and even lead to their merging into the bulk bands without the requirement of closing the band gap or the addition of trivial global perturbations. We have found that there exist resonant points where edge states merge into or emerge from bulk continuum bands and described the possibility of observing these transitions by quenching the strength of an impurity in a topological system. We also envisage technical applications of our method, since the knowledge of the analytical form and properties of edge and bulk modes allow for the reduction of large summations needed for calculating interaction matrix elements to a small number of terms. In large systems and for many-particle problems, such a reduction may be essential in both theoretical and computational applications.


\section*{Acknowledgements}

The authors would like to acknowledge helpful discussions with N. Goldman and M. Di Liberto. C.W.D. acknowledges studentship funding from EPSRC CM-CDT Grant No. EP/L015110/1. P.\"O. and M.V. acknowledge support from EPSRC EP/M024636/1.


%

\end{document}